\documentclass[conference]{IEEEtran}

\usepackage{amsmath,amsfonts}
\usepackage{algorithmic}
\usepackage{array}
\usepackage{textcomp}
\usepackage{stfloats}
\usepackage{url}
\usepackage{verbatim}
\usepackage{graphicx}
\usepackage{balance}

\usepackage{tikz}
\usetikzlibrary{arrows.meta,chains, positioning,shapes.geometric}
\usepackage{subcaption}
\usepackage{makecell}
\usepackage[inline]{enumitem}
\usepackage[font=footnotesize]{caption}
\usetikzlibrary {calc}
\usepackage{romannum}
\usepackage{scalerel}
\usepackage{multirow}
\usepackage{threeparttable}
\usepackage{cite}
\usepackage{diagbox}

\tikzstyle{start} =[ellipse, minimum height=0.001cm, minimum width=0.001cm, text centered, draw=black, fill=yellow, align=center, font=\footnotesize, inner sep=1.5pt]
\tikzstyle{arrow} =[thick,->,>=stealth]
\tikzstyle{process1} =[rectangle, minimum width=0.05cm, minimum height=0.03cm, text centered, draw=black, fill=green!]
\tikzstyle{process_est} =[rectangle, minimum width=0.03cm, minimum height=0.02cm, text centered, draw=black, fill=gray!40]
\tikzstyle{process2} =[rectangle, minimum width=0.05cm, minimum height=0.03cm, text centered, draw=black, fill=cyan!]
\tikzstyle{process3} =[rectangle, minimum width=0.05cm, minimum height=0.03cm, text centered, draw=black, fill=orange!]
\tikzstyle{process4} =[rectangle, minimum width=0.05cm, minimum height=0.03cm, text centered, draw=black, fill=pink!]
\tikzstyle{output} =[rectangle, minimum width=0.05cm, minimum height=0.03cm, text centered, draw=white, fill=white!]
\tikzstyle{compute} =[circle, minimum size =0.001cm, text centered, draw=black, fill=white!]

\begin{document}
\title{Time-based vs. Fingerprinting-based Positioning Using Artificial Neural Networks}

\author{\IEEEauthorblockN{Anil Kirmaz*$^\dagger$, Taylan \c{S}ahin*, Diomidis S. Michalopoulos*, and Wolfgang Gerstacker$^\dagger$}
\IEEEauthorblockA{*Nokia, Munich, Germany \\
$^\dagger$Institute for Digital Communications,
Friedrich-Alexander-Universität Erlangen-Nürnberg,
Erlangen, Germany}
}


\maketitle

\begin{abstract}
High-accuracy positioning has gained significant interest for many use-cases across various domains such as industrial internet of things (IIoT), healthcare and entertainment. Radio frequency (RF) measurements are widely utilized for user localization. However, challenging radio conditions such as non-line-of-sight (NLOS) and multipath propagation can deteriorate the positioning accuracy. Machine learning (ML)-based estimators have been proposed to overcome these challenges. \par 

RF measurements can be utilized for positioning in multiple ways resulting in time-based, angle-based and fingerprinting-based methods. Different methods, however, impose different implementation requirements to the system, and may perform differently in terms of accuracy for a given setting. In this paper, we use artificial neural networks (ANNs) to realize time-of-arrival (ToA)-based and channel impulse response (CIR) fingerprinting-based positioning. We compare their performance for different indoor environments based on real-world ultra-wideband (UWB) measurements. We first show that using ML techniques helps to improve the estimation accuracy compared to conventional techniques for time-based positioning. When comparing time-based and fingerprinting schemes using ANNs, we show that the favorable method in terms of positioning accuracy is different for different environments, where the accuracy is affected not only by the radio propagation conditions but also the density and distribution of reference user locations used for fingerprinting.
\end{abstract}


\section{Introduction}

Position information is essential for a wide range of applications spanning from the industrial to the entertainment sector. Some of the applications, such as  autonomous driving,  process automation and industrial asset tracking require location estimation with high accuracy as emphasized in 3GPP~\cite{3GPPaccuracy}. 

Radio frequency (RF) measurements have been widely utilized for positioning due to their promise for a wide coverage and high accuracy especially when a high bandwidth such as in ultra-wideband (UWB) transmission is available. RF-based positioning can be conducted by various approaches including time-based, angle-based and fingerprinting-based techniques. Here, time-based positioning techniques rely on detecting the time at which the transmitted radio signals arrive at the receiver, i.e., time-of-arrival (ToA). Practical time-based positioning systems require either a clock synchronization among anchor nodes to utilize time-difference-of-arrival (TDoA) measurements or adopt a two-way-ranging (TWR) scheme. On the other hand, fingerprinting techniques rely on a mapping between reference user locations and the corresponding RF measurements utilizing previously collected measurement data. Angle-based techniques perform an estimation of the geometry based on the direction of travel of RF signals, requiring antenna arrays for the direction estimation. In this work, we consider time-based and fingerprinting-based schemes for positioning. 
\par

\subsection{Prior Art}
To attain high accuracy in time-based positioning systems, accurate T(D)oA estimates are required. However, obtaining such accurate estimates is challenging due to  non-line-of-sight (NLOS) or multipath propagation. NLOS propagation occurs when the direct, i.e., line-of-sight (LOS), path between transmitter and receiver is blocked. Such propagation decouples the time-based measurements with the real range, resulting in ranging error. Due to the difficulty in modeling the complex nature of the radio propagation, machine learning (ML) techniques have been used to identify or mitigate this error. \par

Ranging error mitigation has been proposed based on various features extracted from the received waveforms and processed by employing support vector machines (SVMs), Gaussian processes \cite{Xiao,Marano,Kong} and fuzzy comprehensive evaluation\cite{Yu}. Processing channel impulse responses (CIRs) directly with ANNs for ranging error mitigation has been proposed in \cite{Bregar,Angarano, Kirmaz2}, where a significant improvement in ranging or positioning accuracy can be achieved. ANNs have been also proposed to estimate ToA based on a CIR and were shown to outperform various conventional, i.e., non-ML, estimators~\cite{Feigl}. However, the ANN estimator in~\cite{Feigl} was trained mainly with simulation data since obtaining ground-truth ToAs in real-world environments is challenging. In this work, we consider a TWR scheme, which does not require any synchronization between anchor and target device~\cite{McCrady}, thus making the data collection in real world easier. \par 

RF fingerprinting has been widely analyzed in the literature. \textcolor{black}{In the offline phase of fingerprinting, RF measurements referred to as \textit{fingerprints} are collected from known, i.e., reference, user locations. In the online phase, new RF measurements are mapped to a user location based on the relation between the fingerprints and reference user locations obtained via the offline measurements.} Received signal strength (RSS) fingerprinting has been studied as a simple yet efficient alternative for indoor positioning~\cite{RSS_fingeprinting_survey}. However, it is challenging to achieve a high positioning accuracy with RSS fingerprinting since RSS values reflect only a coarse information, with a loose correlation with distance, fluctuating in a dynamic environment. Therefore, \textcolor{black}{fingerprinting based on fine-grained RF measurements such as CSI and }CIR fingerprinting has been advocated for high accuracy positioning since such measurements carry more detailed information related to the radio channel~\cite{fifs,RSS_fingerp}. \textcolor{black}{The mapping between user locations and the corresponding RF fingerprints required for positioning can be accomplished in several ways. A commonly used method is to find the closest, i.e., the most similar, reference fingerprints to the online RF measurements for estimation of user location. However, this approach yields a high positioning complexity for fingerprinting based on high-dimensional entities such as CIRs due to a required search algorithm among a large number of high dimensional CIRs.} Therefore, we employ ANNs to map the CIRs to user locations, similar to~\cite{Niitsoo}. \par

Recently, 3GPP has been studying time-based and fingerprinting-based positioning using ML techniques~\cite{3gpp_one_step}. However, the performance evaluations are done mainly with simulation data based on a standardized channel model.

\begin{figure*}[!t]
\centering
\begin{tikzpicture}[node distance=1.5cm]

\def\factor{0.25}
\def\factors{0.4}
\def\bending{8}

\node (anchor_i) at (-4,4) {\includegraphics[width=.07\textwidth]{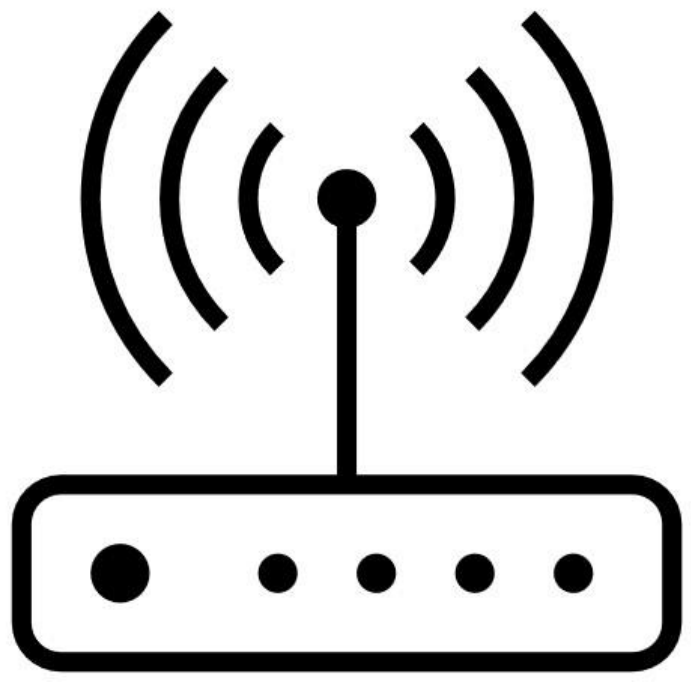}};
\node (anchor_i_name)[output, above of=anchor_i, yshift=-0.65cm] {\footnotesize $\text{Anchor}_2$};

\node (anchor_i_name_btw)[above of=anchor_i_name, xshift= 1.5 cm, yshift=-1.7cm] {\Huge ...};

\node (anchor_j)[right of=anchor_i, xshift=1.4cm, yshift=1cm] {\includegraphics[width=.07\textwidth]{anchor_icon.pdf}};
\node (anchor_j_name)[output, above of=anchor_j, yshift=-0.65cm] {\footnotesize  $\text{Anchor}_n$};

\node (anchor_k)[below of=anchor_i, xshift=+1.2cm, yshift=-0.2cm] {\includegraphics[width=.07\textwidth]{anchor_icon.pdf}};
\node (anchor_k_name)[output, above of=anchor_k, yshift=-0.65cm] {\footnotesize  $\text{Anchor}_1$};

\node (target) at ($(anchor_k) + (2.5,+1.2)$) {\includegraphics[width=.06\textwidth]{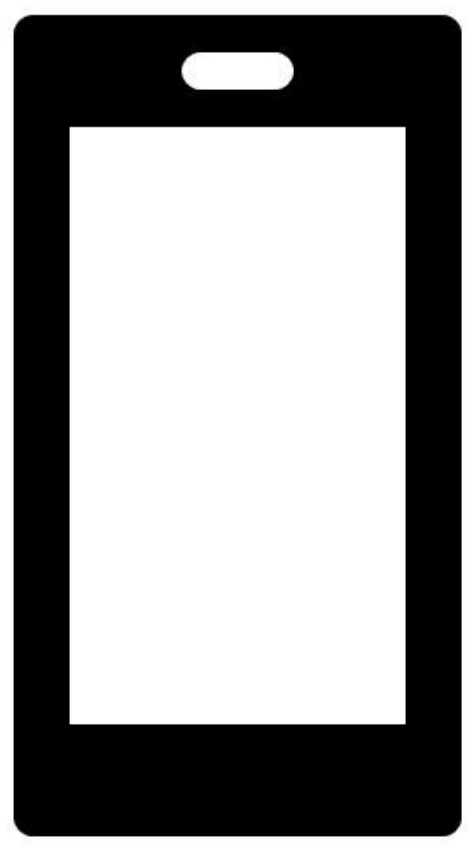}};

\node (target_name)[output, below of=target, yshift=1cm] {\footnotesize \makecell{Target\\device}};

\draw[densely dashed,-{Stealth[length=1.2mm, width=1.5mm]}] ($(anchor_i) - \factor*(anchor_i) + \factor*(target) + (-0.3,0.2) $)  to[bend left=\bending] ($(target) - \factor*(target) + \factor*(anchor_i) + (0.45,0.1) $);

\draw[densely dashed,-{Stealth[length=1.2mm, width=1.5mm]}] ($(target) - \factor*(target) + \factor*(anchor_i)  + (0.45,0.1) $) to[bend left=\bending]  ($(anchor_i) - \factor*(anchor_i) + \factor*(target)  + (-0.3,0.2) $);

\draw[densely dashed,-{Stealth[length=1.2mm, width=1.5mm]}] ($(anchor_j) - 0.5*\factor*(anchor_j) + 0.5*\factor*(target) $)  to[bend left=\bending] ($(target) - \factors*(target) + \factors*(anchor_j) $);

\draw[densely dashed,-{Stealth[length=1.2mm, width=1.5mm]}] ($(target) - \factors*(target) + \factors*(anchor_j) $) to[bend left=\bending]  ($(anchor_j) - 0.5*\factor*(anchor_j) + 0.5*\factor*(target) $) ;

\draw[densely dashed,-{Stealth[length=1.2mm, width=1.5mm]}] ($(anchor_k) - \factor*(anchor_k) + \factor*(target) - (0.15,0.15) $)  to[bend left=\bending] ($(target) - \factor*(target) + \factor*(anchor_k) + (0.2,0.2) $);

\draw[densely dashed,-{Stealth[length=1.2mm, width=1.5mm]}] ($(target) - \factor*(target) + \factor*(anchor_k) + (0.2,0.2) $) to[bend left=\bending]  ($(anchor_k) - \factor*(anchor_k) + \factor*(target) - (0.15,0.15) $) ;

\node (cir_fig)[right of=target, xshift=0.1 cm, yshift=0.2cm] 
{\includegraphics[width=.1\textwidth]{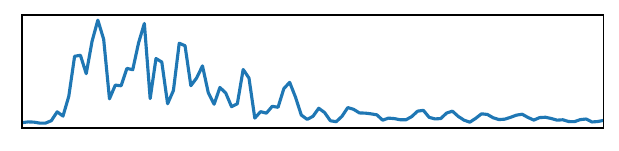}};

\node (CIR_name)[output, above of=cir_fig, yshift=-0.95cm] {\footnotesize $\text{CIR}_1$};

\draw[-] (CIR_name) to ($(CIR_name) + (0,1)$ );

\node (cir_fig2)[right of=target, xshift= 2 cm, yshift=0.2cm] 
{\includegraphics[width=.1\textwidth]{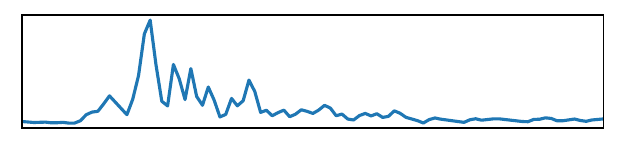}};

\node (CIR_name2)[output, above of=cir_fig2, yshift=-0.95cm] {\footnotesize $\text{CIR}_2$};

\draw[-] (CIR_name2) to ($(CIR_name2) + (0,1)$ );

\node (CIR_name_btw)[right of=target, xshift= 3.5 cm, yshift=0.2cm] {\Huge ...};

\node (cir_fig3)[right of=target, xshift= 5 cm, yshift=0.2cm] 
{\includegraphics[width=.1\textwidth]{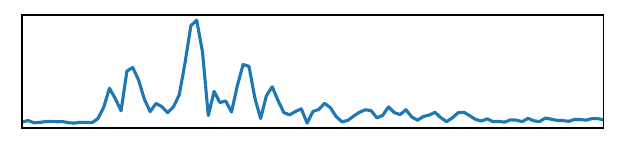}};

\node (CIR_name3)[output, above of=cir_fig3, yshift=-0.95cm] {\footnotesize $\text{CIR}_n$};

\draw[-] (CIR_name3) to ($(CIR_name3) + (0,1)$ );

\node (ann_fingerprint)[process_est, above of=CIR_name3, xshift=1.9cm, yshift=-0.5cm] {\small \makecell{ Fingerprinting}};

\draw[-stealth] ($(CIR_name) + (0,1)$ ) to (ann_fingerprint);

\node (ann_fingerprint_result)[output, right of=ann_fingerprint , xshift=1cm ] {\normalsize $[\widehat{x}, \widehat{y}]$\textsubscript{fingerprinting} };

\draw[-stealth] (ann_fingerprint) to (ann_fingerprint_result);

\node (toa_est)[process_est, below of=cir_fig, yshift=+0.3cm] {\scriptsize \makecell{ ToA \\ estimator}};

\node (toa_est2)[process_est, below of=cir_fig2, yshift=+0.3cm] {\scriptsize \makecell{ ToA \\ estimator}};

\node (toa_est_btw)[below of=CIR_name_btw,  yshift=+0.3cm] {\Huge ...};

\node (toa_est3)[process_est, below of=cir_fig3, yshift=+0.3cm] {\scriptsize \makecell{ ToA \\ estimator}};

\node (toa_out)[output, below of=toa_est, yshift=+0.3cm] {\footnotesize $\widehat{\text{ToA}}_1$};

\draw[-] (toa_out) to ($(toa_out) - (0,1)$ );

\node (toa_out2)[output, below of=toa_est2, yshift=+0.3cm] {\footnotesize $\widehat{\text{ToA}}_2$};

\draw[-] (toa_out2) to ($(toa_out2) - (0,1)$ );

\node (toa_out_btw)[output, below of=toa_est_btw, yshift=+0.3cm] {\Huge ...};

\node (toa_out3)[output, below of=toa_est3, yshift=+0.3cm] {\footnotesize $\widehat{\text{ToA}}_n$};

\draw[-] (toa_out3) to ($(toa_out3) - (0,1)$ );

\node (iterative)[process_est, below of=toa_out3, xshift=1.9cm, yshift=+0.5cm] {\small \makecell{ToA-based positioning \\ algorithm}};

\node (ToA_result)[output, right of=iterative , xshift=1.15cm ] {\normalsize $[\widehat{x}, \widehat{y}]$\textsubscript{ToA} };

\draw[-stealth] ($(toa_out) - (0,1)$) to (iterative);

\draw[-stealth] (iterative) to (ToA_result);

\draw[-stealth] (cir_fig) to (toa_est);
\draw[-stealth] (cir_fig2) to (toa_est2);
\draw[-stealth] (cir_fig3) to (toa_est3);

\draw[-stealth] (toa_est) to (toa_out);
\draw[-stealth] (toa_est2) to (toa_out2);
\draw[-stealth] (toa_est3) to (toa_out3);

\end{tikzpicture}

  \caption{System architecture for ToA-based and fingerprinting-based positioning.}
  \label{fig1}
\end{figure*}

\subsection{Contributions}
To the authors' knowledge, a comparison of ToA-based vs. fingerprinting-based positioning using ANNs based on real-world UWB measurements has not been addressed yet in the literature. The main contributions of this paper are as follows:
\begin{itemize}
    \item We compare the performance of ANN estimators for fingerprinting-based and ToA-based positioning, respectively, adopting a real-world UWB dataset.
    \item We investigate the effect of indoor environments with different propagation characteristics, and density and distribution of reference user locations used for fingerprinting on the system performance.
\end{itemize}

The evaluations in this paper demonstrate the following:
\begin{itemize}
\item The proposed ANN-based ToA estimation yields a better positioning accuracy compared to conventional methods in many cases, and a similar performance in the worst case.

\item The favorable choice between ToA-based and fingerprinting-based positioning using ANNs for a given environment and dataset depends not only on the radio propagation conditions but also the density and distribution of the reference user locations available for fingerprinting.

\end{itemize}
\par

The remainder of this paper is organized as follows. Section~\Romannum{2} describes the considered system model and baseline methods. The proposed schemes, utilized dataset and associated pre-processing are discussed in Section~\Romannum{3}. Section~\Romannum{4} presents numerical performance results, and Section~\Romannum{5} concludes the paper.

\section{System Description}
In this work, we study ToA-based and fingerprinting-based positioning methods based on CIRs. The considered system architecture is depicted in Fig.~\ref{fig1}.

\subsection{CIR}
A CIR characterizes the corresponding communication channel and contains information on the travel time of radio signals from transmitter to receiver. Transmitted signals might arrive at the receiver from different paths, e.g., direct, reflected, or diffracted paths. A CIR of a discrete multipath channel can be expressed with respect to time $t$ and time delay $\tau$ as
\begin{equation}  
\label{CirDef}
\text{CIR}(t,\tau)= \sum_{i=1}^{L} a_i(t) \chi_i(\tau - \tau_i(t)).
\end{equation}
Here, $L$ denotes the number of multipath components, where the $i^{\text{th}}$ multipath component has complex amplitude $a_i(t)$ and delay $\tau_i(t)$, respectively. $\chi_i(\cdot)$ stands for the distorted pulse related to the $i^{\text{th}}$ multipath component, deviating from the ideal Dirac-delta function due to various factors including the finite system bandwidth and frequency dependency of the channel, diffuse scattering and diffraction, and filtering at transmitter and receiver side~\cite{Molisch}. In the remainder of the paper, we will consider CIRs for a fixed $t$ and varying $\tau$ for ToA estimation and CIR fingerprinting. The magnitude of $\text{CIR}(t, \tau)$ for a fixed $t$ and varying $\tau$ will be represented by a function CIR, where $\tau$ and $t$ will be dropped for the sake of simplicity. \par

\subsection{ToA Estimation}
ToA represents the arrival time of the first arriving signal component at the receiver. ToA estimation based on CIR can be formulated as
\begin{equation}
\label{ToaDef}
\widehat{\text{ToA}} = f(\text{CIR}),
\end{equation}
where $f(\cdot)$ and $\widehat{\cdot}$ stand for the function realized by the estimator and the estimated quantity, respectively.

\subsubsection{Conventional ToA Estimators}
In this work, we consider widely used conventional ToA estimation techniques, namely the peak detection and leading-edge detection estimators, which we denote as Peak and LDE, respectively:
\begin{itemize}
    \item \emph{Peak}: The delay time of the first peak of the CIR above a noise threshold is considered as ToA \cite{Falsi}.
    \item \emph{LDE}: A moving average filter is applied to CIR, and the output is further passed through two moving maximum filters of different sizes in parallel. The first delay time, where the output of the smaller moving maximum filter is larger than a noise threshold and exceeds the output of the larger moving maximum filter by a factor, i.e., the leading-edge detection factor, is determined as ToA \cite{Merkl}.
\end{itemize}

We define the noise threshold of a CIR for Peak and LDE estimators in terms of the relative path strength similar to \cite{Lee}, formulated as
\begin{equation}
\label{NoiseDef}
\gamma_{th} = \beta\: \text{max}\{\text{CIR}\}, 
\end{equation}
with the noise threshold factor $\beta$. LDE requires four additional parameters, namely the leading-edge detection factor and the window sizes of the moving average and the two moving maximum filters, respectively. We have optimized the parameters of Peak and LDE by an exhaustive search to yield the lowest mean absolute ToA estimation error.

\subsubsection{Positioning Using ToA Estimates}
Following the ToA estimation described in Section~\Romannum{2}-B and above, we consider the following two algorithms for positioning:
\begin{itemize}
\item \textit{Algo1}: A non-iterative linear least squares approach, similar to~\cite{Bregar}.
\item \textit{Algo2}: A linearized iterative algorithm starting from an initial approximate location~\cite{Pos_algo}.
\end{itemize}
We first use Algo1 and utilize its estimation as the initial approximate location for Algo2. For the sake of completeness, we will compare Algo1 and Algo2 in Section~\Romannum{4}-B.

\subsection{Positioning Using CIR Fingerprints}
We consider CIR fingerprinting, formulated as
\begin{equation}
\label{FingerprintDef}
[\widehat{x}_i, \widehat{y}_i] = g(\text{CIR}_1, \text{CIR}_2, ..., \text{CIR}_L).
\end{equation}
Here, $\widehat{x_i}$ and $\widehat{y_i}$ represent the \textit{x}- and \textit{y}-coordinates of the user location, respectively, estimated based on CIRs from $L$ anchors by using the mapping function $g(\cdot)$. \par

\section{Proposed Positioning Methods}
\subsection{ToA Estimation Using ANNs}

We utilize ANNs for ToA estimation since it is challenging to analytically model the complex nature of NLOS or multipath propagation. A CIR is fed as input to the ANN estimator, ANN\textsubscript{ToA}, for ToA estimation, similar to \cite{Feigl}. We utilize one-dimensional convolutional and residual layers in the ANN to capture local correlations among CIR samples  \cite{LeCun} and to prevent the accuracy degradation problem \cite{He}, respectively. We use two parallel convolutional blocks with depths of one and four, respectively, whose superimposed outputs are input to a cascade of a further convolutional and a fully connected layer. We set the kernel size to 5, i.e., 5x1, in the convolutional layers, with stride of 1. No pooling layer is used in the ANNs in order to avoid a potential information loss. We employ 16 filters, i.e., output channels, in each convolutional layer and the rectified linear unit (ReLU) as activation function. ANN\textsubscript{ToA} is trained based on the Adam optimizer \cite{Kingma} with a batch size of 32 and the mean-squared error (MSE) as cost function. The learning rate is set to $10^{-3}$, and is decreased when the learning converges. The maximum number of training epochs is chosen as 250, with an early stopping mechanism that terminates training once the validation performance of the model does not improve anymore. \par

\subsection{Fingerprinting-based Positioning Using ANNs}
We use ANN\textsubscript{FP} for fingerprinting-based positioning, which has a similar structure as ANN\textsubscript{ToA}. However, as opposed to ANN\textsubscript{ToA}, multiple CIRs are input to ANN\textsubscript{FP}, where each CIR is assigned to an input channel. We set the number of convolutional filters to 32. The output layer comprises two outputs corresponding to the 2D coordinates of the user device, again opposed to ANN\textsubscript{ToA} having a single output (i.e., ToA).

\begin{figure*}[!t]
  \begin{subfigure}{.5\textwidth}
    \centering
    \includegraphics[width = .84\textwidth]{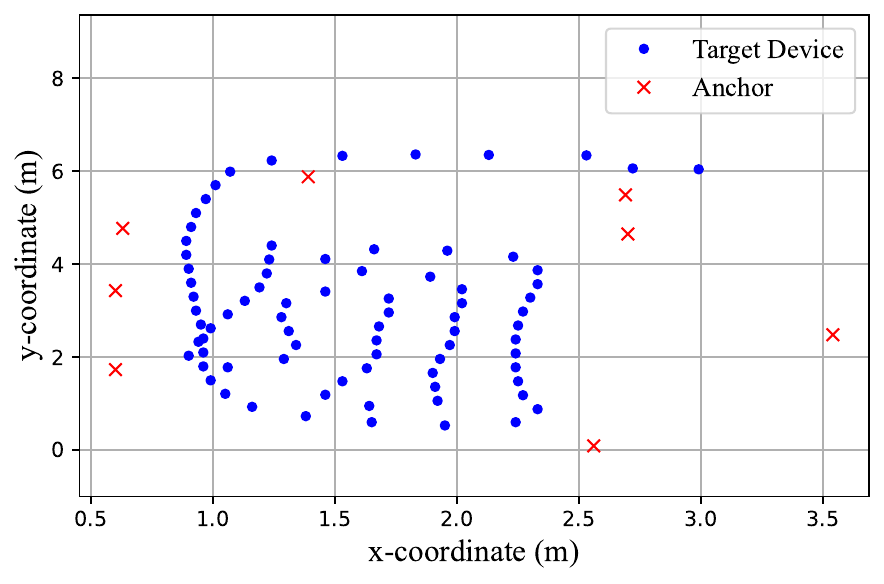}
    \vspace{-0.15cm}
    \caption{}
    \label{fig11a}
  \end{subfigure}
    \begin{subfigure}{.5\textwidth}
    \centering
    \includegraphics[width = .84\textwidth]{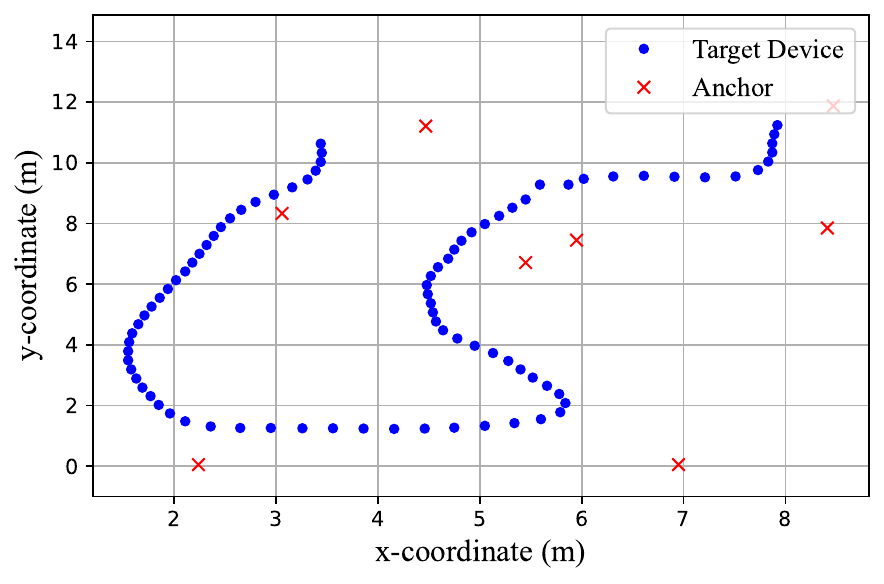}
        \vspace{-0.15cm}
    \caption{}
        \label{fig11b}
  \end{subfigure}

  \vspace{0.3cm}
      \begin{subfigure}{.5\textwidth}
    \centering
    \includegraphics[width = .84\textwidth]{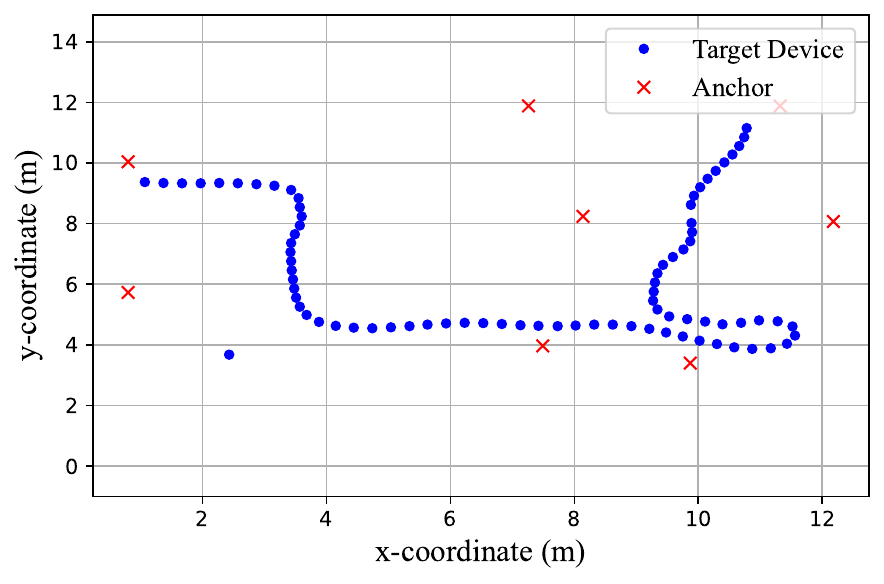}
        \vspace{-0.15cm}
    \caption{}
        \label{fig11b}
  \end{subfigure}
      \begin{subfigure}{.5\textwidth}
    \centering
    \includegraphics[width = .84\textwidth]{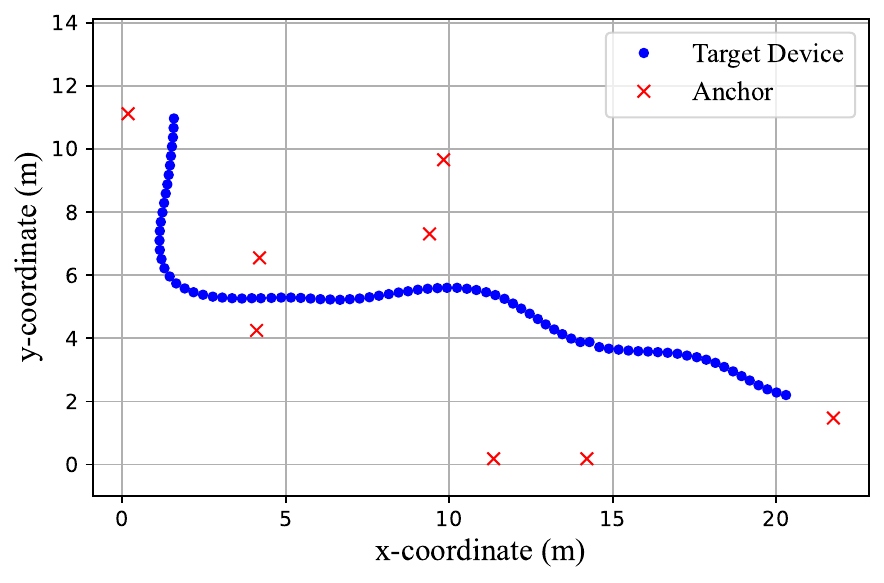}
        \vspace{-0.15cm}
    \caption{}
        \label{fig11b}
  \end{subfigure}
  \caption{2D anchor and user locations in (a) apartment, (b) house, (c) office, and (d) industrial environments.}
      \label{anchor_tag_locations}
\end{figure*}

\subsection{Dataset Description and Pre-processing}
\subsubsection{Dataset}
We have used the publicly available dataset in~\cite{dataset} comprising real-world UWB measurements based on TWR, taken with DWM1000 device~\cite{Decawave}. The dataset contains measurements in four different environments, namely \textit{apartment}, \textit{house}, \textit{office} and \textit{industrial} differing in terms of LOS/NLOS conditions. There are 8 anchors communicating with the target device. The anchor and tag locations in the dataset for different environments are visualized in Fig.~\ref{anchor_tag_locations}. \textcolor{black}{The CIRs are measured for $\sim$80 user locations in each environment, and the measurements are repeated $\sim$30 times. This yields 19200 CIRs with 640 unique anchor-tag links concerning the ToA estimation, and 2400 sets of CIR fingerprints with 80 unique user locations for fingerprinting per environment.}

The measurements were taken with a bandwidth of 499.2 MHz at a center frequency of 3494.4 MHz. The time resolution of the CIRs $\Delta_t$ is $\sim$1 ns \cite{Decawave}. Along with the CIRs, the dataset includes the ToA in terms of the first path delay time and the range estimated by the DWM1000 device as well as the corresponding ground-truth anchor and tag locations. \par

\subsubsection{ToA Labeling}
We determine the ground-truth ToAs, i.e., ToA labels, required to train ANN\textsubscript{ToA} for ToA estimation, according to
\begin{equation}
\label{ToaLabel}
\text{ToA}_i = \widehat{\text{ToA}}_{{\text{DWM}}_i} - \frac{\epsilon_{R_i}}{c\:\Delta_t}.
\end{equation}
Here, $\text{ToA}_i$, $\widehat{\text{ToA}}_{{\text{DWM}}_i}$ and $\epsilon_{R_i}$ stand for the ground-truth ToA, the ToA estimated by DWM1000 and the ranging error, respectively, associated with the $i^{\text{th}}$ CIR. As such, the ranging error is converted into a ToA error and subtracted from the ToA estimated by DWM1000 in order to determine the ground-truth ToA. 

\subsubsection{Data Pre-processing}
For each CIR, only 152 (out of 1016) samples after the first detected path were considered in~\cite{Bregar} since the remaining noise-like samples do not carry significant information about the propagation environment. In addition to these samples, we further consider 10 samples prior to the first detected path in each CIR to compensate for potential first path estimation errors. Each CIR is normalized by its maximum value to avoid a potential bias due to a varying maximum amplitude of the CIRs. \par

\section{Performance Evaluation}
In this section, we present performance results for the ToA-based and fingerprinting-based positioning schemes utilizing the proposed ANN estimators, and the baseline methods described in Sections~\Romannum{3} and \Romannum{2}-B, respectively, based on real-world measurements. We utilize the Pytorch framework \cite{Paszke} to train and test the ANN estimators. \par

The dataset is divided into training, validation and test data with quotas of 70\%, 15\% and 15\%, respectively, for the ANN estimators. To enable a fair comparison, the training and validation data is used to optimize the parameters of the conventional ToA estimators. Training and testing are repeated 10 times with different random selections of the measurement data to average out potential variations across different data chunks. \par

\begin{table}[t]
\centering
\caption{90th percentile ranging error (in cm) of the considered ToA-based estimators in four environments.}
  \vspace{0.1 cm}
\begin{tabular}{ |c|c|c|c|c|c|}
    \cline{2-5}
    \multicolumn{1}{c|}{} & Apartment & House & Office & Industrial   \\
   \hline
    Peak & 75 & 139 & 168 & 232 \\
       \hline
    LDE & 52 & 71 & 109 & 147 \\
       \hline
    ANN\textsubscript{ToA} & 41 & 66 & 108 & 121 \\
   \hline
 \end{tabular}
  \vspace{-0.2 cm}
          \label{ranging}
 \end{table}
 
As the evaluation metric, we consider the 2-norm positioning error, given by
\begin{equation}
\label{RangingError}
\epsilon_{|P_i|} = || [\widehat{x}_i, \widehat{y}_i] - [x_i, y_i] ||_2 ,
\end{equation}
where $\epsilon_{|P_i|}$, $x_i$ and $y_i$ represent the positioning error, and the true \textit{x}- and \textit{y}- coordinates of the location of user $i$, respectively. 

\subsection{Ranging Accuracy Evaluation}
Table~\ref{ranging} shows the 90th percentile ranging error of the considered ToA estimators in four different environments. ANN\textsubscript{ToA} achieves the best ranging accuracy, outperforming LDE by 1-21\%, depending on the environment. The gain of ANN\textsubscript{ToA}
rises to 36-52\% when compared to the Peak estimator. The relatively poor accuracy of Peak can be explained by the susceptibility of the peak detection to multipath propagation \cite{Merkl,Kuhn}. As a further remark, the industrial environment is observed to be the most challenging environment for ToA estimation, followed by office, house and apartment environments, respectively. \par

\begin{figure*}[!t]
  \begin{subfigure}{.5\textwidth}
    \centering
    \includegraphics[width = .9\textwidth]{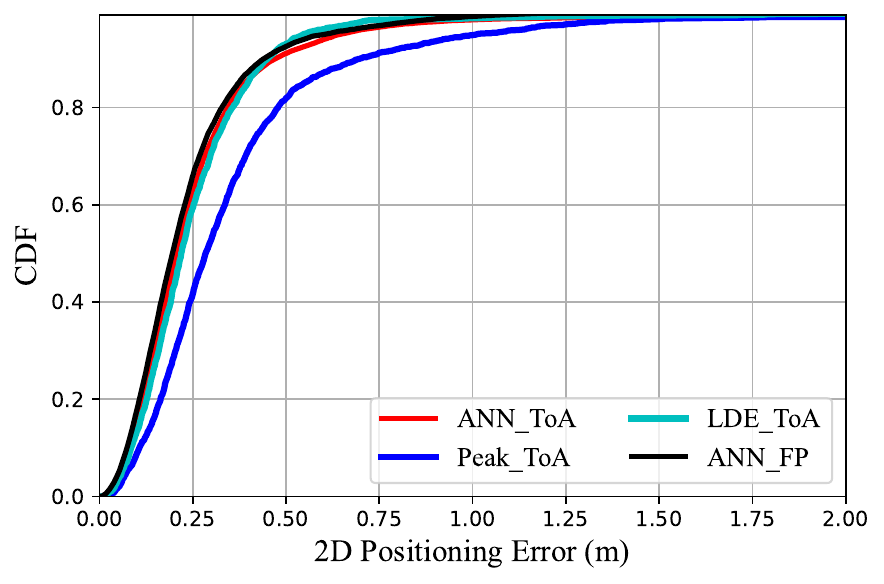}
    \vspace{-0.15cm}
    \caption{}
    \label{fig11a}
  \end{subfigure}
    \begin{subfigure}{.5\textwidth}
    \centering
    \includegraphics[width = .9\textwidth]{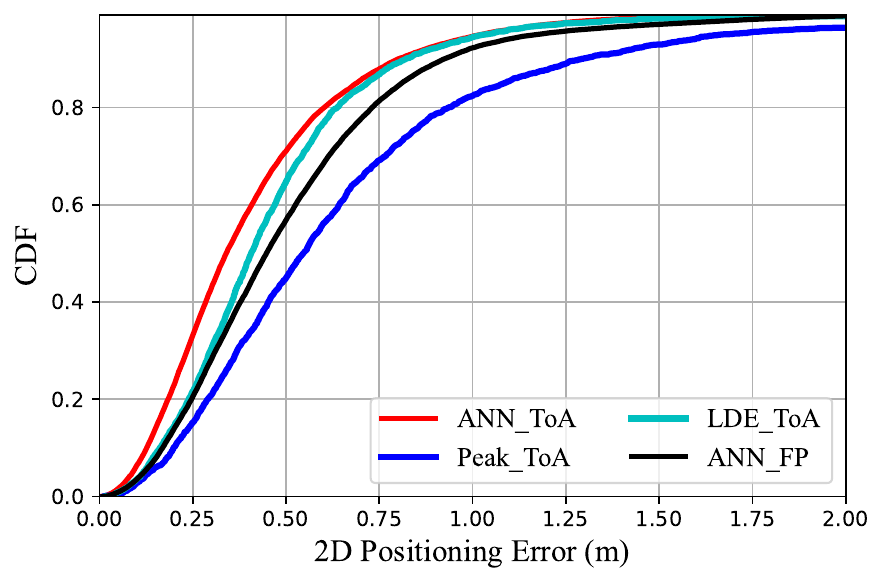}
        \vspace{-0.15cm}
    \caption{}
        \label{fig11b}
  \end{subfigure}

  \vspace{0.3cm}
      \begin{subfigure}{.5\textwidth}
    \centering
    \includegraphics[width = .9\textwidth]{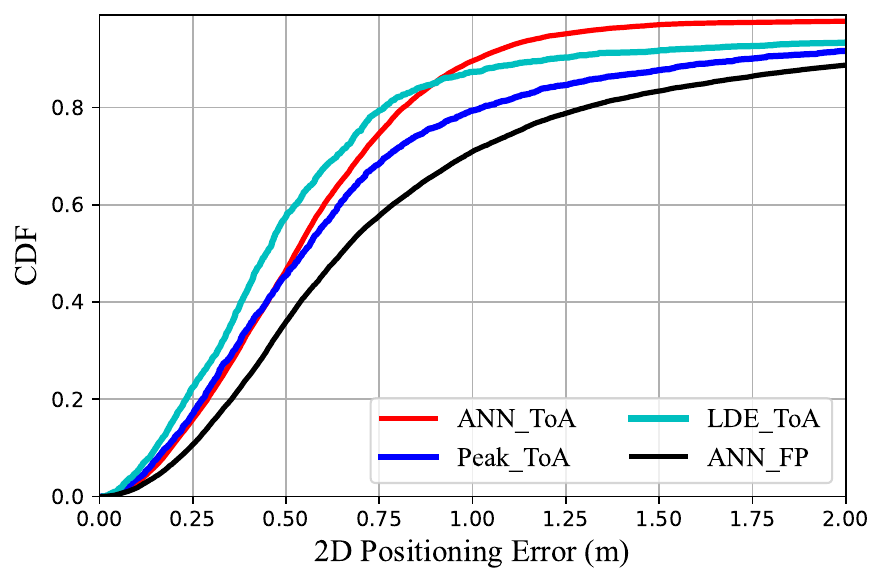}
        \vspace{-0.15cm}
    \caption{}
        \label{fig11b}
  \end{subfigure}
      \begin{subfigure}{.5\textwidth}
    \centering
    \includegraphics[width = .9\textwidth]{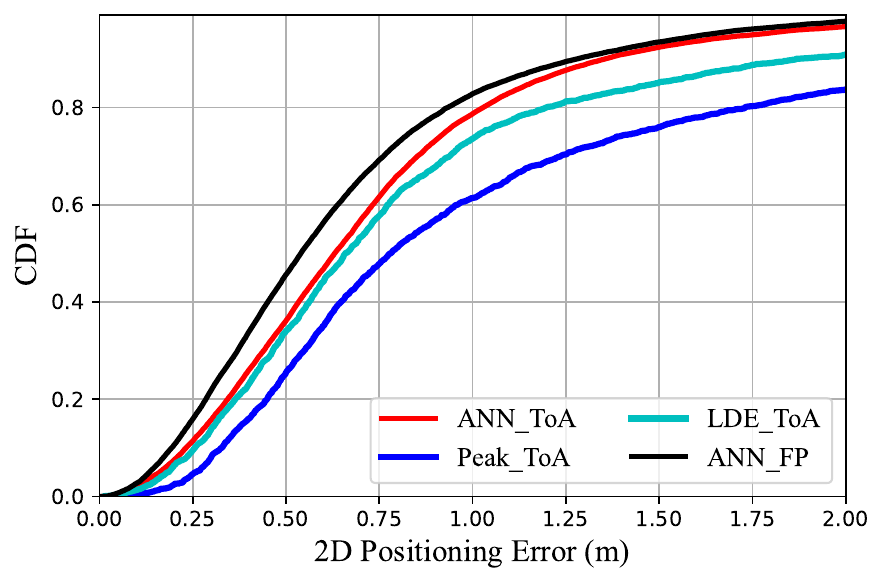}
        \vspace{-0.15cm}
    \caption{}
        \label{fig11b}
  \end{subfigure}
  \caption{CDFs of 2D positioning error of ANN\textsubscript{FP}, ANN\textsubscript{ToA}, LDE and Peak in (a) apartment, (b) house, (c) office, and (d) industrial environments, respectively.}
      \label{positioning_err}
\end{figure*}

\subsection{Positioning Accuracy Evaluation}
Figures~\ref{positioning_err}a-d depict the 2D positioning performance of the considered schemes in the four environments, respectively. ANN\textsubscript{ToA} outperforms LDE by 28\% and 17\% with respect to the 90th percentile in industrial and office environments, respectively. In the house environment, ANN\textsubscript{ToA} yields a lower mean absolute positioning error compared to LDE, whereas the two schemes perform similar in the apartment environment. Peak yields the lowest positioning accuracy, resulting from its poor ranging performance. All ToA-based positioning systems yield the best and worst positioning accuracy in the apartment and industrial environment, respectively, in line with their ranging performance described in Section~\Romannum{4}-A.

An interesting observation is that ANN\textsubscript{FP} and ANN\textsubscript{ToA} yield similar accuracy in the apartment environment, whereas the former yields a worse positioning accuracy in the office and house environments. In other words, the degradation in positioning performance of the fingerprinting approach is more severe than that of the ToA-based scheme when the environment is changed from apartment to the more NLOS dominated house and office environments. This result might seem counter-intuitive as the fingerprinting scheme is considered more robust than the ToA-based scheme to NLOS propagation. Indeed, for the fingerprinting scheme, the accuracy is more related to the quality of the mapping between reference user locations and RF fingerprints. As such, it is desirable to have i) a high density of RF fingerprints (i.e., a high number of RF fingerprints per area); and ii) the reference user locations distributed uniformly in the area of interest to fully capture the relation between the locations and fingerprints. However, the density of RF fingerprints in the house and office environments are lower than that of the apartment environment due to their larger area despite of the similar number of RF fingerprints. Additionally, it is observed from Figs.~\ref{anchor_tag_locations}a-c that the house and office environments are associated with less favorable trajectory-like reference user location distributions compared to the apartment environment with a better distribution. These factors might explain the worse performance of fingerprinting-based positioning in those environments. In the industrial environment, the fingerprinting approach yields a higher positioning accuracy than the ToA-based approach due to the severe performance degradation of the latter. \textcolor{black}{As a further remark, the relatively small dataset size might have had a worse impact on the performance of ANN\textsubscript{FP} compared to ANN\textsubscript{ToA} in all four environments since the dataset size would result in even a smaller number of training data samples for the fingerprinting approach as described in Section~\Romannum{3}-C1.}

\begin{table}[t]
\centering
\caption{90th percentile 2D positioning error (in cm) of different ToA-based positioning algorithms.}
  \vspace{0.1 cm}
\begin{tabular}{ |c|c|c|c|c|c|}
    \cline{2-5}
    \multicolumn{1}{c|}{} & Apartment & House & Office & Industrial   \\
   \hline
    Algo1 & 47 & 88 & 124 & 170 \\
       \hline
    \makecell{Algo2 \\ Initiate @ Closest anchor} & 57 & 107 & 130 & 252 \\
       \hline
    \makecell{Algo2 \\ Initiate @ Algo1 output}  & 47 & 80 & 101 & 136 \\
   \hline
 \end{tabular}
  \vspace{-0.2 cm}
  \label{pos_algos}
 \end{table}

As a result, while the ANN-based schemes can offer superior performance compared to conventional ranging and positioning methods, the choice of using a fingerprinting or time-based positioning scheme based on ANNs would depend on the considered environment, impacting the radio propagation conditions, as well as the density and distribution of the reference user locations for the available training dataset. \par 

We also provide evaluation results for the algorithms Algo1 and Algo2 described in Section~\Romannum{2}-B2 for estimation of the user position from the estimated ToAs. Table~\ref{pos_algos} shows that the highest positioning accuracy is obtained with Algo2 initialized with the location given by the estimate of Algo1, and, therefore, is used to obtain the results in Fig.~\ref{positioning_err} as mentioned in Section~\Romannum{2}-B2.

\section{Conclusions}
In this paper, we have investigated and compared the performance of ToA-based and fingerprinting-based positioning using ANNs, and several conventional positioning techniques. Using real-world measurements, we have observed that the ANN-based ToA estimation yields more accurate positioning compared to conventional ToA estimators in both LOS and NLOS dominated environments with gains of up to 46\% in the 90th percentile positioning accuracy. \par

Comparing ToA-based and fingerprinting-based positioning using ANNs, we have further shown that one of both methods should be selected carefully to achieve a high positioning accuracy taking into account not only the radio propagation conditions but also the density and distribution of reference user locations used for fingerprinting. Further investigations should be conducted to obtain a selection method for a suitable choice of a positioning scheme. Additionally, ANN-based direct TDoA estimation might be investigated in future research for the case that the anchors are synchronized.

\end{document}